\documentclass[preprint,showpacs,aps]{revtex4}

\usepackage{graphicx}
\usepackage{dcolumn}
\usepackage{bm}

\begin{document}

\begin{titlepage}

\title{Quantum Manifestations of Graphene Edge Stress and Edge Instability: A First-Principles Study}

\author{Bing Huang$^1$, Miao Liu$^2$, Ninghai Su$^2$, Jian Wu$^1$, Wenhui Duan$^1$, Bing-lin Gu$^1$, and Feng Liu$^2$\footnote{Email: fliu@eng.utah.edu}}

\address{$^1$Department of Physics, Tsinghua University, Beijing 100084, China}
\address{$^2$Department of Materials Science and Engineering, University of Utah, Salt Lake City, Utah 84112}

\date{\today}

\begin{abstract}
We have performed first-principles calculations of graphene edge
stresses, which display two interesting quantum manifestations
absent from the classical interpretation: the armchair edge stress
oscillates with nanoribbon width and the zigzag edge stress is
noticeably reduced by spin polarization. Such quantum stress effects
in turn manifest in mechanical edge twisting and warping
instability, showing features not to be captured by empirical
potentials or continuum theory. Edge adsorption of H and Stone-Wales
reconstruction are shown to provide alternative mechanisms in
relieving the edge compression and hence to stabilize the planar
edge structure.
\end{abstract}


\maketitle

 \draft

\vspace{2mm}

\end{titlepage}

Graphene, a two-dimensional (2D) single layer of carbon atoms, has
attracted tremendous attention because of its unique electronic
properties\cite{A. H. Castro Neto} and potential applications in
electronic devices\cite{A. K. Geim}. Earlier studies have focused on
characterizing the unusual electronic and transport  properties of
graphene, particularly as a massless Dirac fermion system\cite{A. H.
Castro Neto, A. K. Geim}. More recently, some attention has been
shifted to the structural stability of graphene\cite{A. K. Geim, J.
C. Meyer, A. Fasolino, M. S. Gass}, which is critically important to
realizing the potential applications of graphene. On the one hand,
as a 2D membrane structure, graphene provides an ideal testing
ground\cite{J. C. Meyer, A. Fasolino} for the classical
Mermin-Wagner theorem on the existence of long-range crystalline
order in 2D\cite{Laudau, Mermin}. On the other hand, the free edges
of graphene are amenable to edge instabilities\cite{M. S. Gass, P.
Koskinen, T. Wassmann, V. B. Shenoy}.

The graphene edge stability is characterized by two fundamental
thermodynamic quantities: edge energy and edge stress. The edge of a
2D structure can be understood in analogy to the surface of a 3D
structure\cite{Feng1,Raj}: the edge (surface) $\emph{energy}$
accounting for the energy cost to create an edge (surface) defines
the edge (surface) $\emph{chemical}$ stability; the edge (surface)
$\emph{stress}$ accounting for the energy cost to deform an edge
(surface) defines the edge (surface) $\emph{mechanical}$ stability.
First-principles calculations showed that chemically the armchair
edge is more stable with a lower energy, while the zigzag edge is
metastable against reconstruction\cite{P.  Koskinen}, and both edges
are hydrogenated in an H-rich environment\cite{T. Wassmann}.
Empirical-potential calculations showed that both intrinsic edges
are under compressive stress rendering a mechanical edge twisting
and warping instability\cite{V. B. Shenoy}.

Usually, stress and mechanical instability are understood as
phenomena of classical mechanics, but they are expected to be
affected by quantum effects which become prominent at nanoscale. So
far, however, quantum effects have been mostly shown for electronic
structure and energetic quantities of low-dimensional
nanostructures. Here, we demonstrate an interesting example of
quantum manifestations of mechanical quantities in graphene edge
stress. Using first-principles calculations, we predict that the
armchair edge stress in a nanoribbon exhibits a large oscillation
with ribbon width arising from quantum size effect, while the zigzag
edge stress is reduced by spin polarization. Such quantum effects on
edge stress in turn manifest in graphene edge mechanical
instability, with "quantum" features that apparently cannot be
described by empirical force-field potentials or continuum theory.

%
%

Our calculations were performed using the method based on density
functional theory (DFT) in the generalized gradient approximation,
with the Perdew-Burke-Ernzerhof functional for electron exchange and
correlation potetnials, as implemented in the VASP code \cite{VASP}.
The electron-ion interaction was described by the projector
augmented wave method, and the energy cutoff was set to 500 eV. The
structures were fully optimized using the conjugate gradient
algorithm until the residual atomic forces to be smaller than 10 meV
/\AA. The supercell method with periodic boundary conditions was
adopted to model the graphene nanoribbons (GNR), with a vacuum layer
larger than 15 \AA~ to guarantee that there was no interaction
between the GNR images in the neighboring cells. The edge energy is
calculated as $E_{edge}=(E_{ribbon}-E_{atom})/2L$, where
$E_{ribbon}$ is the total energy of the graphene nanoribbon in the
supercell, $E_{atom}$ is the energy per atom in a perfect graphene
without edge, and $L$ is the length of edge. The edge stress is
calculated as $\sigma_{edge}=\sigma_{xx}/2L$, where $\sigma_{xx}$ is
the diagonal component of supercell stress tensor in the
$x$-direction (defined along the edge), which is calculated using
the Nielsen-Martin algorithm \cite {Nielsen}. All other components
of stress tensor vanish. We also note that DFT is suitable for
calculating ground-state properties of lattice energies and
stresses, to which the non-local many-body effects are not expected
to be important.

\begin{figure}[tbp]
\includegraphics[width=8.0cm]{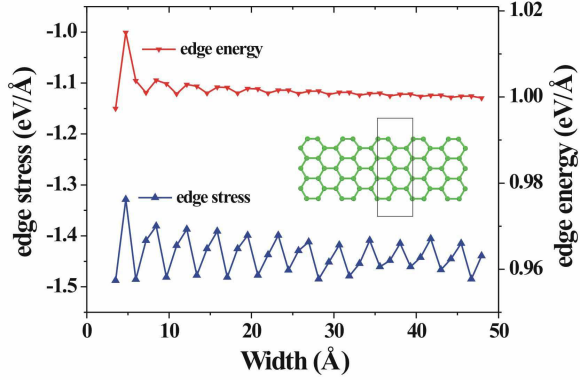}
\caption{ The armchair edge stresses and edge energies of graphene
nanoribbons as a function of ribbon width. Inset: schematics of the
nanoribbon; the rectangle marks one unit cell (supercell) of the
ribbon.}
\end{figure}

Figure 1 shows the calculated edge energy and edge stress of GNR
armchair edges as a function of ribbon width ranging from $\sim$ 3.5
to $\sim$ 48 \AA. One notices that both edge energy and edge stress
oscillate with the increasing width having a period of 3 but out of
phase with each other. The oscillations are originated from the
quantum confinement effect, as also manifested in the similar
oscillations of electron band structures\cite{Nakada, Waka, Son,
Qimin}. The oscillation magnitude of edge energy decays quickly with
the increasing width and converges to $\sim$ 1.0 eV/\AA, which
agrees well with the previous first-principles values\cite{P.
Koskinen}. In contrast, the oscillation magnitude of edge stress
decays much slower with a mean value of $\sim$ -1.45 eV/\AA (using
negative sign as convention for compressive stress). The much larger
oscillation in edge stress than in edge energy is possibly caused by
the fact that edge stress equals to the derivative of edge energy
with respect to strain, so that stress is much more sensitive to the
width-dependent quantum confinement effect. There is also a slight
revival effect in the stress oscillations at $\sim 40\AA$  width,
whose origin is not clear and needs further study.

Figure 2 shows the calculated edge energy and edge stress of GNR
zigzag edges as a function of ribbon width ranging from $\sim$ 5.0
to $\sim$ 85 \AA.  In this case, both edge energy and edge stress
show very weak width dependence and converges quickly, again
consistent with their corresponding electronic-structure
behavior\cite{Nakada, Waka, Son, Qimin}. However, the zigzag edge is
known to have an antiferromagnetic (AFM) ground state\cite{Son}. The
AFM edge energy is calculated to be $\sim$ 1.2 eV/\AA, about 0.2
eV/\AA~ lower than the paramagnetic (PM) edge energy\cite{P.
Koskinen, T. Wassmann, Hosik Lee}. Thus, we have calculated the edge
stress of both AFM and PM states for comparison. We found that spin
polarization have a sizable effect reducing the compressive stress
from $\sim$ -0.7 eV/\AA~ in the PM edge to $\sim$ -0.5 eV/\AA~ in
the AFM edge.

\begin{figure}[tbp]
\includegraphics[width=8.0cm]{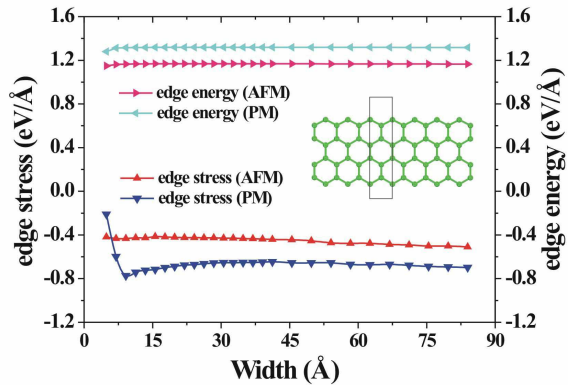}
\caption{ The AFM and PM zigzag edge stresses and edge energies of
graphene nanoribbons as a function of ribbon width. Inset:
schematics of the nanoribbon; the rectangle marks one unit cell
(supercell) of the ribbon.}
\end{figure}

Our first-principles stress calculations confirm qualitatively the
recent empirical-potential results\cite{V. B. Shenoy} that both
edges are under compressive stress. However, there are also some
significant differences. Two quantum manifestations of edge stress
stand out, which are absent from the empirical prediction. One is
the quantum oscillation of armchair edge stress with the increasing
nanoribbon width, and the other is the reduction of zigzag edge
stress by spin polarization. The physical origin of edge energy and
edge stress is associated with the formation of one dangling bond on
each edge atom. The repulsive interaction between the dangling bonds
is believed to be one origin for the 'compressive' edge stress. In
addition, in the armchair edge, it is well-known\cite{T. Kawai} that
the edge dimers form triple \textbf{-C$\equiv$C-} bonds with a much
shorter distance ($\sim$ 1.23 \AA~ according to our calculation)
adding extra compressive stress to the edge; while in the zigzag
edge, spin polarization further reduces the compressive stress.
Therefore, quantitatively, the armchair edge has a much larger
compressive stress ($\sim$ -1.45 eV/\AA ) than the zigzag edge
($\sim$ -0.5 eV/\AA). In contrast, the empirical potentials
predicted a smaller compressive stress in the armchair edge ($\sim$
-1.05 eV/\AA ) than in the zigzag edge ($\sim$ -2.05 eV/\AA)\cite{V.
B. Shenoy}.

The quantum effects in edge stress will in turn modify the
mechanical edge instability. The compressive edge stress means the
edge has a tendency to stretch. If we apply a uniaxial in-plane
strain   to a nanoribbon along the edge direction, the strain energy
can be calculated as \cite{V. B. Shenoy}
\newline
$E_{str}  = 2\tau _e L\varepsilon  + E_e L\varepsilon ^2  +
\frac{1}{2}E_s A\varepsilon ^2$               (1)
\newline

Here, $\emph{A}$ is the ribbon area, $\emph{L}$ is the edge length,
$\tau _e$ is the edge stress, $E_e$ is the 1D edge elastic modulus
in a 2D nanoribbon, in analogy to the 2D surface elastic modulus in
a 3D nanofilm\cite{J. Zang}, and $E_s$ is the 2D sheet elastic
modulus. Since $\tau _e$ is negative, for small enough tensional
strain $\varepsilon$ (positive), the negative first term (linear to
$\varepsilon$) in Eq. (1) can always overcome the positive second
and third terms (quadratic to $\varepsilon$) to make $E_{str}$
negative. Consequently, the ribbon is unstable against a small
amount of stretching along the edge direction. Fitting
first-principles calculations, by manually deforming the sheet and
ribbon along the edge direction, to equation (1), we obtained $E_s$
$\approx$ 21.09 eV/\AA$^{2}$, $E_e$(amchair)$\approx$ 3 eV/\AA~ and
$E_e$(zigzag) $\approx$ 24 eV/\AA~ with $\tau _e$ already calculated
above directly (see Figs. 1 and 2). Our $E_s$ value is in good
agreement with the experiment\cite{C. Lee} and empirical simulation
result\cite{V. B. Shenoy}. But our $E_e$ values are notably
different from the empirical results\cite{V. B. Shenoy}.

Another effective way to stretch the edge of a 2D sheet is by
out-of-plane edge twisting and warping motions, which are
barrierless processes. For example, assuming a sinusoidal edge
warping with displacement $\mu _e  = a\sin (2\pi x/\lambda )$ of
amplitude $\emph{a}$ and wavelength $\lambda$, which decays
exponentially into the sheet as $e^{ - y/l}$ (See inset of Fig. 3),
where $\emph{l}$ is the decay length, Shenoy $\emph{et al.}$ have
shown that minimization of strain energy leads to characteristic
length scales of such warping instability as $l \approx 0.23\lambda$
and $a \approx \sqrt {( - \lambda \tau _e )/(1.37E_b  + 14.8E_e
/\lambda )}$. Using their empirical-potential values of $\tau _e$,
$E_e$ and $E_s$, they estimated that the warping magnitude of
armchair edge is smaller than that of zigzag edge and both are
larger than typical thermal fluctuations\cite{V. B. Shenoy}.

\begin{figure}[tbp]
\includegraphics[width=8.0cm]{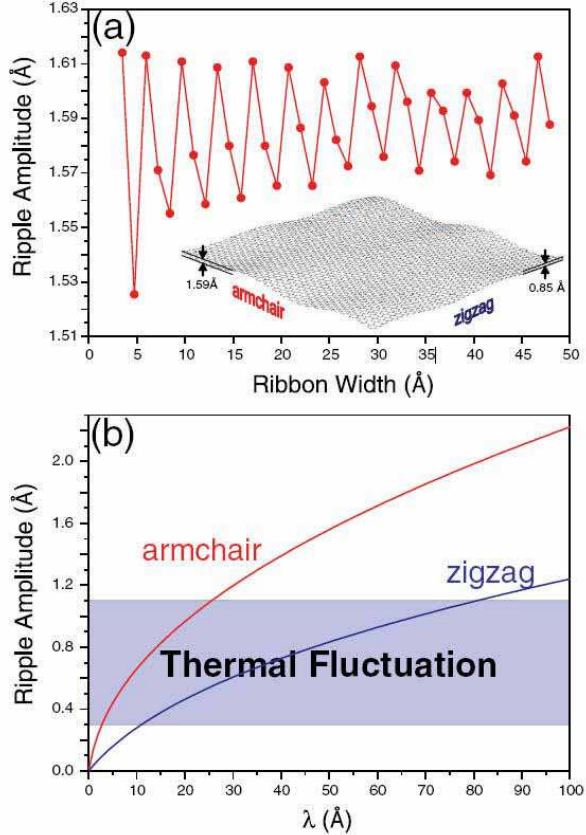}
\caption{ (a) Armchair edge ripple amplitude versus ribbon width for
$\lambda$ = 50\AA. Inset: Schematics of ripple formation along the
armchair and zigzag edge for $\lambda$ = 50\AA. (b) Armchair (red)
and zigzag (blue) edge ripple amplitude as a function of $\lambda$.
Light blue band shows the typical range of thermal fluctuation. }
\end{figure}

Our first-principle predictions, however, are different in several
ways. First, absent from empirical prediction, the quantum
oscillation of $\tau _e$ of armchair edge gives rise to an
oscillating armchair edge warping amplitude for given wavelength as
a function of nanoribbon width, as shown in Fig. 3a. Second,
opposite to empirical prediction, the warping amplitude of armchair
edge is much larger than that of zigzag edge, as shown in Figs. 3b.
Third, the mechanical undulation of zigzag edges induced by
compressive edge stress is comparable to thermal
fluctuations\cite{J. C. Meyer, A. Fasolino}, as shown in Fig. 3b,
and hence the two are difficult to distinguish. We also note that in
addition to the continuum analysis we perform here, the accurate
first-principles values of $\tau _e$ and $E_e$ can be used as input
parameters for the finite element simulations of large graphene
systems\cite{V. B. Shenoy}.

%
%

Because the compressive edge stress is partly originated from the
dangling bond, naturally, we may saturate the dangling bonds to
relieve the compressive stress. We have tested this idea by
saturating the edge with H that indeed confirmed our physical
intuition. For armchair edge in a 1-nm wide ribbon, we found H
saturation changes the edge stress from -1.42 eV/\AA~ to  -0.35
eV/\AA; for zigzag edge in a 2.0-nm wide ribbon, it changes the edge
stress from -0.42 eV/\AA~ to +0.13 eV/\AA. Therefore, the H edge
saturation, or saturation by other molecules in general, is expected
to relieve the edge compression. Especially, it can reverse the
compressive stress in a zigzag edge to tensile.

\begin{figure}[tbp]
\includegraphics[width=8.0cm]{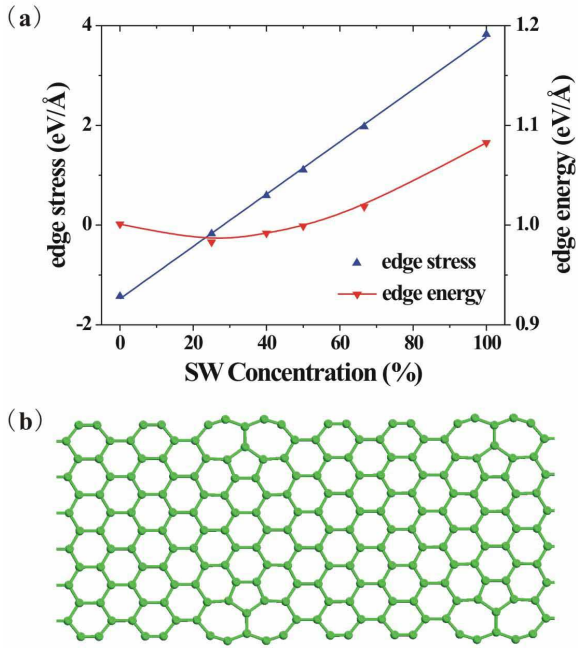}
\caption{ (a) The armchair edge stresses (with linear fit) and edge
energies of graphene nanoribbons as a function of edge SW defect
concentration. (b) The optimaized ribbon structure at the 50\% SW
defect concentration.}
\end{figure}

Surface reconstruction has long been known as an effective mechanism
in relieving surface stress\cite{Feng2}. Thus, in addition to H
saturation, we have investigated possible edge reconstructions in
relieving the edge compressive stress. We have considered the
Stone-Wales (SW) defect\cite{A. J. Stone}, which appealed to us
become a SW defect in 2D is equivalent to a dislocation core in 3D
that is known as a common stress relieve mechanism. Figure 4a shows
the calculated armchair edge stress along with edge energy as a
function of one type of SW defect (7-5-7 ring structure)
concentration. Figure 4b shows an example of the optimized edge
structure at the 50\% defect concentration. One sees from Fig. 4a
that the edge stress increases linearly from compressive to tensile
with the increasing SW defect concentration. The most stable edge
structure is at $\sim$ 25\% defect concentration where the edge
stress is very small and slightly compressive. A small stress value
indicates that this chemically stable edge structure (with the
lowest edge energy) is also most mechanically stable against
deformation.

Figure 5a shows the ground-state AF zigzag edge stress along with
edge energy as a function of another type of SW defect (5-7 ring
structure) concentration. Figure 5b shows an example of the
optimized edge structure and spin charge density at the 50\% defect
concentration. Again, the edge stress increases linearly from
compressive to tensile with the increasing defect concentration, the
same as the case of armchair edge (Fig. 4a), but the edge energy
decreases monotonically with the most stable edge having 100\% of
defects, in agreement with a recent first-principles calculation
\cite{P. Koskinen}. The initial compressive edge stress ($\sim$ -0.5
eV/\AA) is completely reversed to a large tensile value of $\sim$1.2
eV/\AA~ in the most stable edge. Also, the 100\% defected edge
becomes non-spin-polarized. In general, the zigzag edge spin
decreases continuously with the increasing SW defect concentration,
similar to the behavior found previously for other types of defects
\cite{Bing}.

\begin{figure}[tbp]
\includegraphics[width=8.0cm]{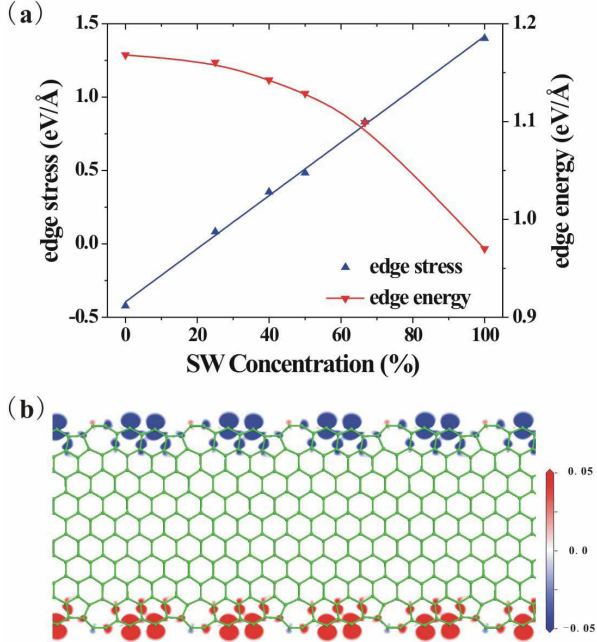}
\caption{ (a) The zigzag edge stresses (with linear fit) and edge
energies of graphene nanoribbons as a function of SW defect
concentration. (b) The optimized ribbon structure and spatial
distribution of spin density (charge density difference between
spin-up and spin-down states in units of $\mu_{B}$ \AA$^{-2}$) of
the AFM ground state at the 50\% SW defect concentration. }
\end{figure}

We note that there is a first-principles calculation of graphene
edge stress was reported recently\cite{S. Jun}, but that calculation
appears inconsistent with all other existing first-principles
calculations . Their edge energies differ from all previous
results\cite{P. Koskinen, T. Wassmann} and ours that are consistent
with each other. The well-known spin-polarization of the zigzag edge
was not considered, the quantum oscillation of armchair edge energy
and stress was not shown, and their edge stress were not calculated
from direct quantum-mechanical formula \cite{Nielsen}.

In conclusion, quantum effects have been widely shown for electronic
structure and energetic quantities of low-dimensional
nanostructures. We demonstrate, in addition, quantum manifestations
of mechanical quantities in graphene edge stress.  We show that
quantum confinement can lead to stress oscillations and spin
polarization can reduce stress, which in turn "quantum mechanically"
modify the edge twisting and warping instability. We further show
that H edge saturation and SW edge reconstruction can not only
improve the 'chemical' stability of graphene edges by lowering the
edge energy as shown before\cite{P. Koskinen, T. Wassmann}, but also
enhance their 'mechanical' stability by converting compressive edge
stress towards tensile and hence stabilizing the planar edge
structure. Our first-principles findings, which cannot be captured
by classical methods, provide new insights to the understanding of
mechanical stability of graphene. We suggest that experimental
measurement of armchair edge ripple magnitudes for different widths
and comparison with zigzag edge ripples could confirm our
predictions of quantum effects in graphene edge stresses
differentiating from empirical results. We expect the quantum
manifestation of mechanical properties such as stress to exist
generally in many low-dimensional nanostructures.

The work at Tsinghua is supported by the Ministry of Science and
Technology of China and NSFC, the work at Utah is supported by
DOE.

\newpage

\end{document}